\documentclass[12pt]{article}
\usepackage[utf8]{inputenc}
\usepackage{heck}
\usepackage{amsmath,amssymb}
\usepackage{color}
\definecolor{myblue}{rgb}{.97,.97,1}
\usepackage{CJK}
\usepackage{hyperref}
\usepackage{url}
\usepackage{graphicx}  
\usepackage{cite}
\usepackage{mathrsfs}

\usepackage{ascmac}
\usepackage{fancybox}

\begin{document}


\begin{CJK*}{UTF8}{min}

\title{Comments on Short Multiplets \\ in Superconformal Algebras\footnote{To appear in proceedings for the Pollica Summer Workshop on "Mathematical and Geometric Tools for Conformal Field Theories'', June 3-21, 2019.}}

\authors{\centerline{Masahito Yamazaki (山崎雅人)}}

\institution{IPMU}{\centerline{Kavli IPMU (WPI), University of Tokyo, Kashiwa, Chiba 277-8583, Japan}}

\date{October, 2019}

\abstract{
The problem of classifying all short multiplets of 
superconformal algebras still seems to be an open question. 
A generic short multiplet is non-unitary, which nevertheless is of interest in various contexts.
Even if one is interested in unitarity theories only,
non-unitary short multiplets are of use in the analysis of (super)conformal blocks.
The classification problem is mathematically formulated in terms of the representation theory of parabolic Verma modules, whose theory is known to be more challenging than that of 
 more standard Verma modules associated with the Borel subalgebra. We comment on some recent developments of the representation theory, which could be of help in solving the classification problem.}


\maketitle 
\end{CJK*}


\section{The Problem}

On the occasion of the Pollica Summer Workshop ``Mathematical and Geometric Tools for Conformal Field Theories'', the organizers have asked me to write up a very short note on an open question or a research direction. I have taken advantage of this opportunity to highlight the following problem, which seems to be little appreciated by the community. It is indeed surprising that the following problem is still unsolved:\footnote{The problem posed here seems to have been solved for the particular case of four-dimensional $\mathcal{N}=1$ supersymmetry
in a nice paper by Li and Stergiou \cite{Li:2014gpa}. The author would like to thank Andreas Stergiou for 
bring this paper to author's attention.}

\bigskip
\begin{itembox}[n]{Classification Problem}
Classify all possible short multiplets (i.e.\ highest-weight irreducible representations) of (super)conformal algebras, in general ($D\ge 3$) spacetime dimensions. 
\end{itembox}

\section{Isn't This Well-Known Already?}

The most likely reaction to the above-mentioned problem is that the answer to this problem should have been known since decades ago. I myself have long thought that this should be the case. However, almost all of the existing literature on the representation theory of (super)conformal algebras (e.g.\ \cite{Evans,Mack:1975je,Flato:1983te,Dobrev:1985qv,Dobrev:1985vh,Dobrev:1985qz,Minwalla:1997ka,Dobrev:2002dt,Bhattacharya:2008zy,Buican:2016hpb,Cordova:2016emh}) have focused on {\it unitary} representations of the (super)conformal algebras. In general, there are many more {\it non-unitary} irreducible representations, and a generic irreducible representation is non-unitary.\footnote{For example, a free massless scalar field in flat space satisfies the Klein-Gordon Equation $\Box \phi=0$. There is a generalization of this equation to $\Box^{k\ge 2} \phi=0$  (see \cite{Brust:2016gjy} for recent discussion), which describes a non-unitary theory.}

\section{Why Interesting?}

I hope the importance of this question is obvious to readers.  In the case of the two-dimensional conformal algebra (i.e.\ Virasoro algebra), one arrives at the celebrated minimal models by studying irreducible representations \cite{Belavin:1984vu}. These models subsequently have been applied to universality classes for various systems in high energy theory, condensed matter theory and statistical mechanics. It is in my opinion rather natural to ask a similar question in higher dimensions.

While most of the short multiplets are non-unitary, non-unitary theories could still be applied to many problems in physics and mathematics, e.g.\ statistical mechanics. Even if one is interested only in unitary theories, non-unitary representations do appear in the analysis of (super)conformal blocks, which are crucial inputs for the modern bootstrap program \cite{Poland:2018epd}. Namely, we can study the superconformal blocks as an analytic complex function of the scaling dimension of the intermediate operator,
and many of the poles of the block corresponds to non-unitary short multiplets (see the literature on the 
recursion relation for conformal blocks \cite{Kos:2013tga,Penedones:2015aga,Iliesiu:2015akf,Erramilli:2019njx}).

\section{Why Not Verma Module/Kac Criterion?}

One can easily think of a straightforward approach for the problem. Again, we can learn from the study of the two-dimensional Virasoro algebra: we can study a Verma module, which is a universal module for a highest-weight state representation. We can then study when the module is irreducible (with the help of the Kac determinant formula), and we can decompose the 
Verma module into irreducible components.

Indeed, the study of the Verma module of a finite-dimensional semisimple Lie algebra is well-studied subject in mathematics (see e.g.\ the book \cite{Humphreys_BGG} for a survey). 

However, the situation is complicated by the fact that what is relevant for a physics of (super)conformal
algebra is not the ordinary Verma module, but rather its variant, the parabolic Verma module (introduced first in \cite{Lepowsky} in mathematics literature). The latter is associated with a parabolic subalgebra, whereas the former to
 the Borel subalgebra: for this reason let us call an `ordinary' Verma module a `Borel' Verma module. 
 This point concerning parabolic and Borel Verma modules seems to be known to some experts since long ago, except not emphasized enough in the literature; see \cite{Yamazaki:2016vqi} for a recent discussion. 

One might think that the distinction between a Borel Verma module and a parabolic
Verma module is minor. However, it is known in the mathematical literature there is a subtle difference between the representation theories between the two, and a complete understanding of this subtlety is still out of 
reach (see chapter 9 of \cite{Humphreys_BGG} for a good summary and e.g. \cite{Lepowsky,Boe,Matsumoto} for some early references).\footnote{Here is a more technical explanation for mathematically-included readers, for illustration of the subtlety. Let us denote the Borel Verma module with highest weight $\lambda$ by $M(\lambda)$, and its parabolic counterpart, associated with a parabolic subalgebra $\mathfrak{p}$, by $M_{\mathfrak{p}}(\lambda)$ (we here follow the notations of \cite{Yamazaki:2016vqi}). When $M(\lambda)$
is reducible at location $\mu$ there is a homomorphism $M(\mu)\to M(\lambda)$. This naturally induces
a homomorphism $M_{\mathfrak{p}}(\mu)\to M_{\mathfrak{p}}(\lambda)$ (the so-called {\it standard} homomorphism), and hence one is tempted to conclude that $M_{\mathfrak{p}}(\lambda)$ is also reducible. However, it is not guaranteed in general that this morphism is non-zero. It is also the case that not all the homomorphisms of $M_{\mathfrak{p}}(\mu)\to M_{\mathfrak{p}}(\lambda)$ have their counterparts $M(\mu)\to M(\lambda)$; such homomorphisms are called {\it non-standard}. In other words, irreducibility of $M(\lambda)$ is neither sufficient nor necessary for the irreducibility of the corresponding $M_{\mathfrak{p}}(\lambda)$} For this reason the study of representation theory of parabolic Verma modules is still an active area of 
research.\footnote{The irreducible decomposition of the parabolic Verma module is already `known' in the following sense. Suppose that one knows the decomposition of the Borel Verma module $M(\lambda)$ into 
irreducible modules $L(\mu)$:
$M(\lambda) = \sum_{\mu \prec \lambda } c^{\mu}{}_{\lambda} L(\mu)$.
On the other hand, one has the relation for the characters between Borel and parabolic Verma modules (see \cite[Lemma 1]{Jantzen} and \cite[Lemma 2]{Oshima:2016gqy}):
${\rm ch}\, M_{\mathfrak{p}}(\lambda)=\sum_{w\in W_{\mathfrak{l}}} \textrm{det}(w) \, {\rm ch}\, M(w.\lambda)$, 
where $W_{\mathfrak{l}}$ is a Weyl group of a reductive subalgebra $\mathfrak{l}$ defined in \cite{Oshima:2016gqy}.
By combining the two equations, one obtains the 
decomposition of the parabolic Verma module $M_{\mathfrak{p}}$
into irreducible components. However, this approach involves the sum of the Weyl group $W_{\mathfrak{l}}$, and does not seem to be too efficient when the rank of the Weyl group is large. The author would like to thank Hisayoshi Matsumoto for discussion on this point.}

As in the case of the Virasoro algebra,
irreducibility criterion of
parabolic Verma modules is provided by the determinant formula (an analog of the Kac determinant formula for the Virasoro algebra).
The precise expression for the determinant formula was derived long ago by a mathematician Jantzen in 1977 \cite{Jantzen}, however little attention has been paid to his work in the physics literature.
For a superconformal algebra, the relevant determinant formula was conjectured in \cite{Yamazaki:2016vqi}
and was later proven only recently in \cite{Oshima:2016gqy} by Oshima and the author (which generalizes the previous work by Gorelik and Kac \cite{Gorelik}). The general irreducibility criterion of \cite{Oshima:2016gqy}
was analyzed more concretely in \cite{Sen:2018del}.\footnote{Note that irreducibility criterion of \cite{Oshima:2016gqy} for superconformal algebras is different from the the so-called ``Kac criterion'' \cite{Kac_1986}---the former applies to parabolic Verma modules, while the latter to Borel Verma modules. This point is worth emphasizing since several papers on representation theories on superconformal algebras, such as \cite{Dobrev:1985qv,Dobrev:1985vh,Dobrev:1985qz,Minwalla:1997ka}, refer to the Kac criterion, whose applicability
to a physical (super)conformal algebra is not clear for the reason mentioned above. Note that the relevance of 
Jantzen's criterion is certainly known to some mathematicians---for example, the paper \cite{EHW} 
who mathematically classified unitarity irreducible representations of the conformal algebra does refer to the
Jantzen's criterion.}

Of course, working out the irreducibility criterion is only part of the problem,
and one still needs to first find the decompositions into irreducible representations,
and find explicit characterization for each of the irreducible modules. 
Even for non-supersymmetric conformal algebras 
explicit and comprehensive description of the null states is recent \cite{Penedones:2015aga,Erramilli:2019njx},
and one can ask if we can complete the program for general superconformal algebras.
Such a result will also be of great use in writing down a recursion relation for superconformal blocks along the lines of \cite{Penedones:2015aga}.

In summary, despite various developments over the past decades, the classification problem of 
short multiplets of superconformal algebras is still an open problem, either mathematically or physically. Since this is a well-defined question
with many potential applications, I would like to challenge (super)conformal-field-theory and/or representation-theory aficionados to settle this question once and for all. If anyone becomes interested in this problem and starts thinking about it after reading this article, then I could say that this article already served it purpose.

\section*{Acknowledgments}

The author would like to thank Hisayoshi Matsumoto, Yoshiki Oshima and Kallol Sen for discussion
over the past several years.  The author would like thank  2019 Pollica summer workshop, which was supported in part by the Simons Foundation (Simons Collaboration on the Non-perturbative 
Bootstrap) and in part by the INFN. The author is partially supported by WPI program (MEXT, Japan) and by JSPS KAKENHI Grant No.\ 17KK0087, No.\ 19K03820 and No.\ 19H00689.


\bibliographystyle{nb}
\bibliography{bootstrap_Pollica}

\end{document}